\documentclass[12pt]{article}
\usepackage{pdproc,epsfig} 

\begin{document}

\renewcommand{\thefootnote}{\alph{footnote}}

\title{
NEW VIEWS ON THE PROBLEM OF CP VIOLATION~\protect\footnote{
Written version of a talk presented at the 
``Second Workshop on Neutrino Oscillations in Venice'' 
(NOVE-II), Venice, Italy. 3-5th, December 2003.}} 

\author{HISAKAZU MINAKATA}

\address{Department of Physics, Tokyo Metropolitan University, \\
1-1 Minami-Osawa, Hachioji, Tokyo 192-0397, Japan\\
 {\rm E-mail: minakata@phys.metro-u.ac.jp}}

%  \centerline{\footnotesize and}

%\author{SECOND AUTHOR'S NAME}

%\address{Group, Company, Address, City, State ZIP/Zone,Country}

\abstract{
After briefly recollecting basic features of the good-old way of 
detecting CP violation by comparing $\nu_{e}$ and $\bar{\nu}_{e}$ 
appearance measurement in long-baseline (LBL) neutrino oscillation 
experiments, I discuss two new ways of exploring leptonic CP violation. 
First, I discuss the reactor-LBL method in which reactor measurement 
of $\theta_{13}$ is combined with $\nu_{e}$ (no $\bar{\nu}_{e}$) 
appearance measurement in LBL. 
Assuming $\sim 10^{3}$ GW$_{th}$$\cdot$ton$\cdot$year operation 
of a reactor experiment, CP sensitivity at 90 \% CL is shown 
to exists in $\sin^2{2\theta_{13}} > 0.03$ (0.04) with 
2 years running of Hyper-Kamiokande (10 years running of SK). 
Second, I review the method which I call the BNL strategy, 
in which one tries to explore CP violation through observing 
oscillatory pattern of neutrino oscillation.
%I make some comments, some sweet and a sweating ones. 
%
Motivated by the approval of the JPARC neutrino program, I also 
discuss a low-energy realization of the BNL strategy. 
It is meant to measure neutrino oscillation at the first and the 
second oscillation maxima by two HKs placed at Kamioka and 
somewhere in Korea. It is suggested by very rough argument that 
8 years running in $\nu$ and $\bar{\nu}$ modes with one HK at 
Kamioka is equivalent to 2 years running in $\nu$ mode with two 
HK complex.}

\normalsize\baselineskip=15pt

\section{Introduction}

Under the title kindly given to me by Milla, to whom we all would 
like to thank for her enthusiasm of having the workshop in such a place 
of scenic beauty, Venice, I will try to describe some new as well as 
the old ways of detecting CP violation. 
Exploring leptonic CP violation is the right topics in the workshop 
in this location; 
we should remember, in trying to discover CP violation, the 
courage possesed by Marco Polo who was brave enough to go on 
voyage to unexplored world and even have reached close to a far 
distant country Japan!

I think it timely to discuss ways of exploring leptonic CP violation at the 
end of 2003. About a year ago, the first result of KamLAND \cite{KamLAND} 
told us that the solution to the solar neutrino problem is given by the 
MSW large-mixing-angle (LMA) solution \cite{wolfenstein,MSW}. 
It not only beautifully settled the problem which lasted 
nearly 40 years \cite{bahcall} but also opened the door to our 
expedition to a CP violating world.
Now, the two solar neutrino experiments, SNO and Super-Kamiokande (SK), 
both state that the larger $\Delta m^2_{12}$ LMA-II region is excluded 
at 99 \% CL in the analysis with 2 degrees of freedom \cite{sno_salt,SK_dn}.
By combining them the statistical significance of the statement is now 
99.9 \% CL \cite{smy}. Therefore, we have finally pinned down the 
unique parameter region in the (1-2) sector of lepton flavor mixing, 
LMA-I. 
Together with the pioneering discovery of neutrino oscillation in the 
atmospheric neutrino observation by Super-Kamiokande \cite{SKatm}, 
which is followed by the confirmation by K2K \cite{K2K}, 
we now know the structure of lepton flavor mixing in the 
(1-2) and (2-3) sectors of the MNS matrix \cite{MNS}. 
Thus, we are left with the determination of the structure of its (1-3) sector, 
$\theta_{13}$ and CP violating phase $\delta$.

First, I would like to mention the physics motivations for 
detecting leptonic CP violation. 
Why do we want to know about leptonic CP violation? 
My strongest motivation comes from better understanding of the concept 
of lepton-quark correspondence. 
Nowadays we all know from quantum anomaly consideration of the 
standard model that quarks and leptons are related with each other in a 
deeper level; we cannot remove any one of them without ruining 
the theory. 
We should remember, however, that our prejudice based on the experience 
in the quark sector badly failed in the lepton sector in which the large 
mixing angles are quite a commonplace event. I think it important to 
confirm (or refute) our prejudice that the leptonic Kobayashi-Maskawa 
phase is unsuppressed. 
It is truly remarkable that the importance of the concept of lepton-quark 
correspondence was recognized early in sixties by Shoichi Sakata and 
coworkers. They developed the Nagoya model, the neutrino-based 
unified model of quarks and leptons \cite{nagoya}. 

Another strong motivation comes from leptogenesis \cite{FY86}, 
the best candidate mechanism to date for generating baryon number 
asymmetry in the universe. 
Although there is no direct connection between the leptonic 
Kobayashi-Maskawa phase $\delta$ and the phases which are 
responsible for leptogenesis at high-energies, it is conceivable that 
the low-energy phase $\delta$ contains certain combinations of the 
effect of high-energy CP violating phases. 
For more about this possibility, see Pascoli's talk \cite{pascoli} in 
this workshop.

\section{Comparing Neutrino and Antineutrino Appearance Experiments; 
Good-Old Way of Measuring CP Violation}

\subsection{A Brief History}

Let me start by describing the story in good-old days of detecting leptonic 
CP violation. Though it is not the main concern in my talk, let me mention 
about it briefly because it is the right time to recollect the past and 
foresee the future. 
(I must, however, warn the readers that nothing is new in this section.)
To the best of my knowledge, it was first noted by 
Cabibbo \cite{Cabibbo} that leptonic CP violation can be detected 
by observing the difference between neutrino oscillation probabilities 
$P \equiv P(\nu_{\mu} \rightarrow \nu_{e})$ and 
$\bar{P} \equiv P(\bar{\nu}_{\mu} \rightarrow \bar{\nu}_{e})$.
In vacuum, it takes the form 
\begin{eqnarray}
P - \bar{P} \equiv P(\nu_{\mu} \rightarrow \nu_{e}) - 
P(\bar{\nu}_{\mu} \rightarrow \bar{\nu}_{e}) = - 
4 J 
\left(\frac{\Delta m^2_{12} L}{4E}\right)
\sin^2{\left(\frac{\Delta m^2_{13} L}{4E}\right)}, 
\label {PbarPvac}
\end{eqnarray}
under the approximation that 
$\frac{\Delta m^2_{12}}{\Delta m^2_{13}}$ is small, 
where $J$ stands for the Jarlskog factor 
$J = c_{12}s_{12}c_{23}s_{23}c^2_{13}s_{13}\sin{\delta}$. 
Despite the pioneering work done in seventies, we needed long time 
to start discussing seriously how to detect CP violation 
in an experimentally realistic setting. It is partly because 
we did not expect that the Jarlskog factor which controls the 
order of magnitude of the CP violating effect can be as large as 
$J/\sin{\delta} \simeq 0.04$ 
(assuming the saturation of the CHOOZ bound \cite{CHOOZ})  
as we know today.

The modern revival of interests in observing leptonic CP violation was 
triggered by works done in 1996-97 by Arafune {\it et al.} and others 
\cite{arafune,tanimoto,MNprd97,MNplb97}.  
It was then followed by spurts of works on varying subjects which are 
too many to quote in this report. 
For more complete references including the ones in early era, see e.g., 
Refs.~\cite{MNprd97,MNjhep}. 
The topics included the issue of earth matter effect 
which might obscure the effect of CP violating phase, and where is 
the region of large CP violating effects.

People gradually recognized that there are two options, the low- and 
the high-energy options, to detect CP violation and measure $\delta$. 
(For this classification, see \cite{NOW_mina}.)
The low-energy option \cite{yasuda,koike-sato,MNplb00} is 
advantageous because effect of CP violation is large at low energies.  
It is practically the unique option for conventional superbeam type 
experiments whose idea may be traced back to \cite{MNplb00,lowecp}.
It is nice to see that the idea of low-energy superbeam has concrete 
realization as feasible experimental programs; 
Refs.~\cite{SPL}, \cite{JHF}, and \cite{NuMI} from low to high energies.

On the other hand, the high-energy option was very popular among 
people despite that CP violating effects are small. People coined into 
the high-energy option because they favored the idea of neutrino factory 
\cite{geer,RGH,nufact} 
which utilizes intense neutrino flux from muon storage ring, 
whose flux is (supposed to be) so high as to easily overcome the 
smallness of the effects. The advantage of using high energy beam 
in neutrino factory comes from the features that beam convergence and 
cross sections are larger while the background is lower. 
(See Ref.\cite{golden} and Ref.\cite{pinney-yasuda} for most 
complete sensitivity analyses in which detailed experimental 
conditions and full correlations of errors are taken into account, 
respectively.)

The question of low- vs. high-energy options has not been settled yet.  
I personally believe that the low-energy option is the correct 
strategy to explore leptonic CP violation, and certainly it is 
the unique choice for tomorrow. 
But the ultimate answer to the question of which option is more profitable 
depends upon the size of $\theta_{13}$ and remains to be seen.

\subsection{Displaying CP and Matter Effects in Terms of Bi-Probability Plot}

So far so many words on ``history''. Let us now return to physics. 
Measuring CP violation effect by observing $P-\bar{P}$ suffers 
from the problem of matter effect contamination. Fortunately, 
the matter effect is not overwhelming but is comparable with genuine 
CP violation due to the leptonic Kobayashi-Maskawa phase in 
relatively short (among LBL) baseline experiments such as the 
JPARC-SK project \cite{JHF}. 
Then, it would be nice if we have a tool for representing these competing 
effects in a transparent way. In fact there exists a such tool called 
the "bi-probability plot" which was introduced in \cite{MNjhep}.

\begin{figure}[h]
\begin{center}
\epsfig{figure=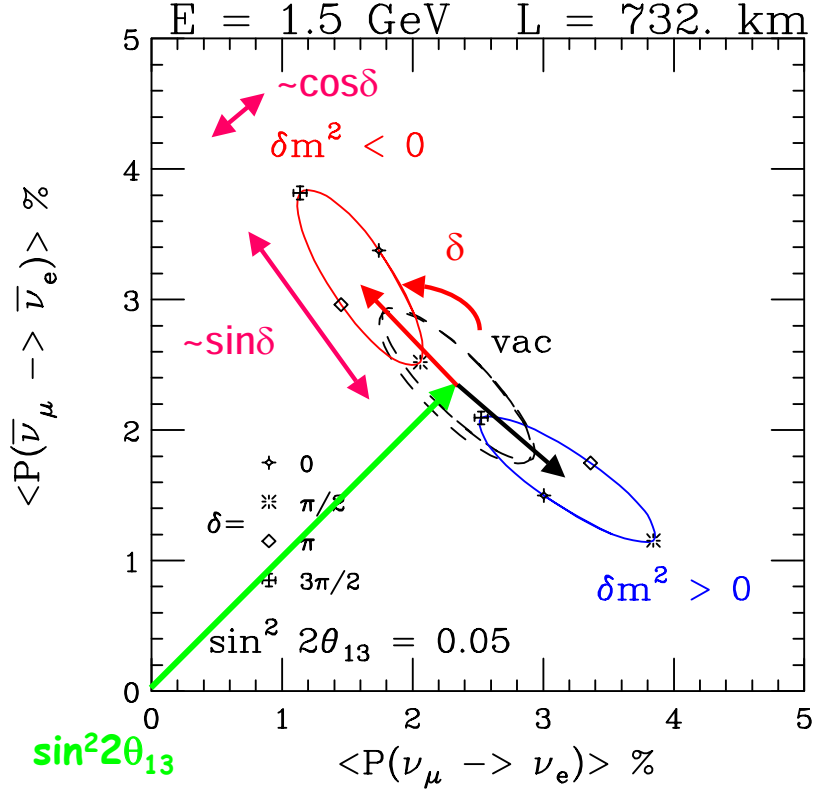,width=11.0cm}
\end{center}
\caption{
A $P$- $\bar{P}$ bi-probability plot with experimental parameters 
corresponding to NuMI off-axis project is presented for the purpose 
of exhibiting characteristic features of the neutrino oscillations 
relevant for low-energy superbeam experiments. 
Namely, it can disply competing three effects, CP violating and CP 
conserving effects due to $\delta$ as well as the matter effects 
in a compact fashion. For more detailed description of its properties, 
see \protect\cite{MNjhep}.
}
\label{fig:biprobability}
\end{figure}

In Fig:~\ref{fig:biprobability} we present a typical example 
which was kindly prepared by 
Adam Para using relevant parameters in NuMI off-axis project for his 
presentation somewhere. 
One can observe from Fig.~1 that the three effects, 
the CP violating and CP conserving effects due to $\delta$ as well as 
the matter effect, are represented in a compact way in a single diagram. 
By giving two observable $P$ and $\bar{P}$, you can draw a dot in 
$P-\bar{P}$ space, and it becomes a closed ellipse when $\delta$ is 
varied.

As indicated in Fig:~\ref{fig:biprobability} 
the lengths of major and minor axes 
(the ``polar'' and ``radial'' thickness of the ellipses) 
represent the size of the $\sin{\delta}$ and the $\cos{\delta}$ 
terms, respectively,  
whereas the distance between two ellipses with positive and negative 
$\Delta m^2_{13}$ displays the size of the matter effect. 
Finally, the distance to the center of the ellipse from the origin is 
essentially given by $\sin^2{2 \theta_{13}}$. 
Notice that all the features of the bi-probability plot except for distance 
between $\Delta m^2_{13}=\pm$ ellipses are essentially determined 
by the vacuum parameters in setting of $E$ and $L$ relevant for 
the superbeam experiments. Therefore, one can easily guess how it looks 
like in the other experimental settings. 
As indicted in Fig:~\ref{fig:biprobability}, 
the CP violating and CP conserving effect of 
$\delta$ are comparable in size with the matter effect.

Utility of the bi-probability plot is not just limited to the 
$P-\bar{P}$ plot as in Fig:~\ref{fig:biprobability}, and it has 
wider applications. It was used to make features of comparison 
of different two experiments and/or measurement more 
transparent \cite{MNP3}. 
We will see more examples of applications in the following sections.

\subsection{$P-\bar{P}$; CP or Matter Effects?}

Now, I just want to make a clarifying remark. 
It is often claimed that one can determine the sign of $\Delta m^2_{13}$ 
by measuring $P-\bar{P}$.  On the other hand, I have said that 
$P-\bar{P}$ tells you the effect of CP violation.  
How these two facts are made consistent with each other? 

Let me answer the question. I first note that 
\begin{eqnarray}
P(\bar{\nu}_{\mu} \rightarrow \bar{\nu}_{e}; \Delta m^2_{13}, \delta, a)  &=& 
P(\nu_{\mu} \rightarrow \nu_{e}; \Delta m^2_{13}, -\delta, -a) 
\nonumber \\ 
&\simeq&  
P(\nu_{\mu} \rightarrow \nu_{e}; -\Delta m^2_{13}, \pi + \delta, a),
\end{eqnarray}
where $a=\sqrt{2} G_F N_e$ is the famous index of refraction 
of Wolfenstein \cite{wolfenstein}. The last equality approximately holds 
\cite{MNP1} due to the fact that $\Delta m^2_{12} \ll \Delta m^2_{13}$.
Then, $\Delta P \equiv P-\bar{P}$ can be written as 
\begin{eqnarray}
\Delta P
&\equiv& 
P(\nu_{\mu} \rightarrow \nu_{e}; \Delta m^2_{13}, \delta, a) - 
P(\bar{\nu}_{\mu} \rightarrow \bar{\nu}_{e}; \Delta m^2_{13}, \delta, a)  
\nonumber \\ 
&\simeq& 
P(\nu_{\mu} \rightarrow \nu_{e}; \Delta m^2_{13}, \delta, a) - 
P(\nu_{\mu} \rightarrow \nu_{e}; -\Delta m^2_{13}, \pi + \delta, a). 
\label {PbarPmatt}
\end{eqnarray}
From (\ref{PbarPmatt}) we can observe the following:

\vskip 0.2cm

\noindent 
(1) If a measurement is done under the environment that dependence on 
CP phase $\delta$ is not sizable 
(which is not difficult to achieve because the effect is small anyway), 
then $\Delta P$ tells you the sign of $\Delta m^2_{13}$. 
If $\Delta P$ is positive, the sign of $\Delta m^2_{13}$ 
is positive and vice versa. 

\vskip 0.2cm

\noindent 
(2) To detect leptonic CP violation as cleanly as possible, 
it is better to go to the region with less matter effect. 
This is the strategy exploited by the low-energy option for measuring 
CP violation \cite{MNplb00}.

\subsection{T Violation; Cleanest Way of Detecting Genuine CP Violating Effect}

Let me mention very briefly that measurement of T Violation is the 
cleanest way of detecting genuine CP violating effect, though it is 
not easy to carry it out experimentally. It is because the oscillation 
probability can be written on very general ground as \cite{KTY} 
\begin{eqnarray}
P(\nu_{\alpha} \rightarrow \nu_{\beta}) =
A \cos{\delta} + B \sin{\delta} + C
\end{eqnarray}
where $A$, $B$, and $C$ are functions of $\Delta m^2_{13}$ and $a$.
Then, the T violating measure $\Delta P_{T}$ is given by 
\begin{eqnarray}
\Delta P_{T} &\equiv&  
P(\nu_{\alpha} \rightarrow \nu_{\beta}; \Delta m^2_{13}, \delta, a) - 
P(\nu_{\beta} \rightarrow \nu_{\alpha}; \Delta m^2_{13}, \delta, a) 
\nonumber \\
&=&
2B \sin{\delta}
\label{DeltaT}
\end{eqnarray} 
for symmetric matter profile. It stems from the fact that 
only the coefficient $B$ is antisymmetric under the interchange 
$\alpha \leftrightarrow \beta$. Therefore, if $\Delta P_{T} \neq 0$, 
then $\delta \neq 0$ even in matter. 
The matter effect cannot create fake T violation. 
(Note, however, that the matter effect modifies the coefficient $B$
in Eq.~(\ref{DeltaT}), whose feature is made transparent in \cite{PW01}.)

\begin{figure}[h]
\begin{center}
\epsfig{figure=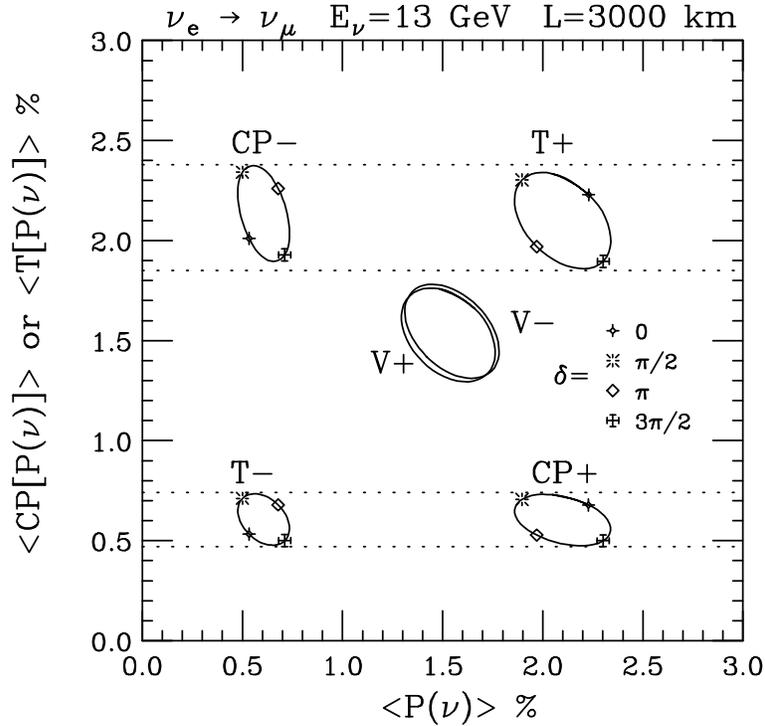,width=10.0cm}
\end{center}
\caption{
A simultaneous $P$- $T[P]$ and $P$- $\bar{P}$ bi-probability plot 
with experimental parameters
corresponding to (nearly) optimal energy and baseline of maximal 
exhancement of T violating effect \protect\cite{PW01}.
Notice the difference between movement of the direction in 
$P$- $T[P]$ and $P$- $\bar{P}$ plot; they are orthogonal with 
each other for reasons explained in the text.
}
\label{fig:PTP}
\end{figure}

Again the feature of T violating measure $\Delta P_{T}$ can be clearly 
represented by the similar bi-probability plot \cite{MNP1}. 
As shown in Fig:~\ref{fig:PTP} the matter effect splits positive and negative 
$\Delta m^2_{13}$ ellipses but along the diagonal in the $P-T[P]$ plot.
It is because the point of $\delta = 0$ on ellipses must remain on the 
diagonal line, as dictated by (\ref{DeltaT}).
On the other hand, the movement of the two ellipses are along orthogonal 
(``polar'') direction in the $P-CP[P]$ bi-probability plot 
as exhibited in Fig:~\ref{fig:PTP}. 
A secret behind the structure of ``baseball diamond''  
in the simultaneous $P-T[P]$ and $P-CP[P]$ bi-probability plot 
(Fig:\ref{fig:PTP}) is explained in \cite{MNP1}.

To carry out the T violation measurement, we must wait for the 
construction of an intense electron (anti-) neutrino beam either by 
beta beam \cite{beta} or neutrino factory.

\section{New Ways of Measuring CP Violation}

We now address the original topics of this talk, new ways of detecting 
CP violation. We discuss two approaches. 
The first one is by combining LBL measurement with reactor experiment, 
a possibility raised quite recently \cite{MS03}.
The second one is the strategy pursuit by people in Brookhaven National 
Laboratory \cite{BNL}. Let me call the latter, the BNL strategy.

\subsection{Reactor-LBL Combined Method}

We start with the reactor-LBL combined method. 
The reasoning behind the proposal is as follows.
In conventional way of detecting CP violation by LBL experiments 
one has to measure appearance probability not only in neutrino 
channel but also in antineutrino channel. But, there are greater 
difficulties in the antineutrino appearance measurement compared 
with that in the neutrino mode. They include a factor of 3 smaller 
reaction cross sections and severer background, even ignoring the issue 
of slightly less intense $\bar{\nu_{\mu}}$ beam. 
The former implies that 3 times longer running time is required for 
accumulation of equal number of events with that in neutrino channel.

Therefore, it would be nice if there exists another way of complementing 
neutrino mode appearance measurement in LBL experiments. 
A natural possibility is the reactor experiment for $\theta_{13}$ 
as a pure measurement of this angle independent of other mixing 
parameters \cite{MSYIS}.

The principle of detection of CP violation by the reactor-LBL combined 
method is in fact very simple. 
LBL $\nu_{e}$ appearance experiment will observe the 
neutrino oscillation probability $P(\nu_{\mu} \rightarrow \nu_{e})$. 
In leading order in $\Delta m^2_{21}/\Delta m^2_{31}$ and $s^2_{13}$,  
it takes the form \cite{golden}
\begin{eqnarray}
P(\nu_{\mu} \rightarrow \nu_{e}) \equiv 
P(\nu)_{\pm} = X_{\pm} s_{13}^2 + 
Y_{\pm} s_{13} \cos {\left( \delta \pm \frac{\Delta_{31}}{2} \right)} + 
P_{\odot},
\label{Pmue}
\end{eqnarray}
where $\pm$ refers to the sign of $\Delta m^2_{31}$. 
The coefficients $X_{\pm}$,  $Y_{\pm}$, and $P_{\odot}$ 
are given by 
\begin{eqnarray}
X_{\pm} &=& 4 s^2_{23} 
\left(
\frac{\Delta_{31}}{B_{\mp}} 
\right)^2
\sin^2{\left(\frac{B_{\mp}}{2}\right)}, 
\label{X} \\
Y_{\pm} &=& \pm 8 c_{12}s_{12}c_{23}s_{23}
\left(
\frac{\Delta_{21}}{aL}
\right)
\left(
\frac{\Delta_{31}}{B_{\mp}}
\right)
\sin{\left(\frac{aL}{2}\right)}
\sin{\left(\frac{B_{\mp}}{2}\right)},
\label{Y}\\
P_{\odot} & = & c^2_{23} \sin^2{2\theta_{12}} 
\left(\frac{\Delta_{21}}{aL}\right)^2
\sin^2{\left(\frac{aL}{2}\right)}
\end{eqnarray}
with 
\begin{eqnarray}
\Delta_{ij}  \equiv \frac{|\Delta m^2_{ij}| L}{2E}
%\quad \quad J_r &\equiv& c_{12}s_{12}c_{23}s_{23}c_{13}^2s_{13},
\quad
{\rm and} \quad B_{\pm} \equiv \Delta_{31} \pm aL,
\end{eqnarray}
where $a = \sqrt{2} G_F N_e$ denotes the index of refraction 
in matter with $G_F$ being the Fermi constant and $N_e$ a constant 
electron number density in the earth. 
We use in this paper the standard notation of the MNS matrix.
The mass squared difference of neutrinos 
is defined as $\Delta m^2_{ji} \equiv m^2_j - m^2_i$ where
$m_i$ is the mass of the $i$-th eigenstate.

Let me simplify the discussion by taking the energy corresponding 
the first oscillation maximum $\Delta_{13} = \pi$. 
In fact, there exist a number of reasons for tuning the beam 
energy to the oscillation maximum in doing the appearance 
and the disappearance measurement in LBL experiments, 
as listed in \cite{KMN02}. In this case, 
$\cos{\left(\delta \pm \frac{\Delta_{13}}{2} \right)} = \mp \sin{\delta}$ 
and (\ref{Pmue}) can be solved for  $\sin{\delta}$ as 
\begin{eqnarray}
\sin{\delta} = 
\frac{P(\nu) - P_{\odot} - X_{\pm} s_{13}^2}
{\mp Y_{\pm} s_{13}}.
\label{sdelta}
\end{eqnarray}
We note that, since $\theta_{13}$ can be measured by reactor 
experiments, the right-hand side (RHS) of (\ref{sdelta}) consists 
solely of experimentally measurable quantities. Therefore, LBL 
measurement of $P(\nu_{\mu} \rightarrow \nu_{e})$, when combined 
with the reactor experiment, implies measurement of $\sin{\delta}$.
One can easily show that the argument can be generalized to the 
case off the oscillation maximum.

The real question is, however, if it is really possible to carry out
such reactor-LBL combined measurement of CP violation 
in a realistic setting. If yes, the next question is 
what would be the required conditions for the reactor
and the LBL experiments for the purpose.
To answer the questions we have performed a detailed statistical
analysis assuming reasonable systematic errors.
For LBL experiment, we take the neutrino-mode appearance 
measurement in the JPARC-HK(SK) experiment which includes 
the effect of background after severe cut for $\pi^{0}$ rejection.
For reactor $\theta_{13}$ experiment, we assume the systematic errors
which are likely to be achieved based on our estimation \cite{MSYIS} 
in designing the reactor experiment at the cite of Kashiwazaki-Kariwa 
nuclear power plant.

\begin{figure}[h]
\begin{center}
\epsfig{figure=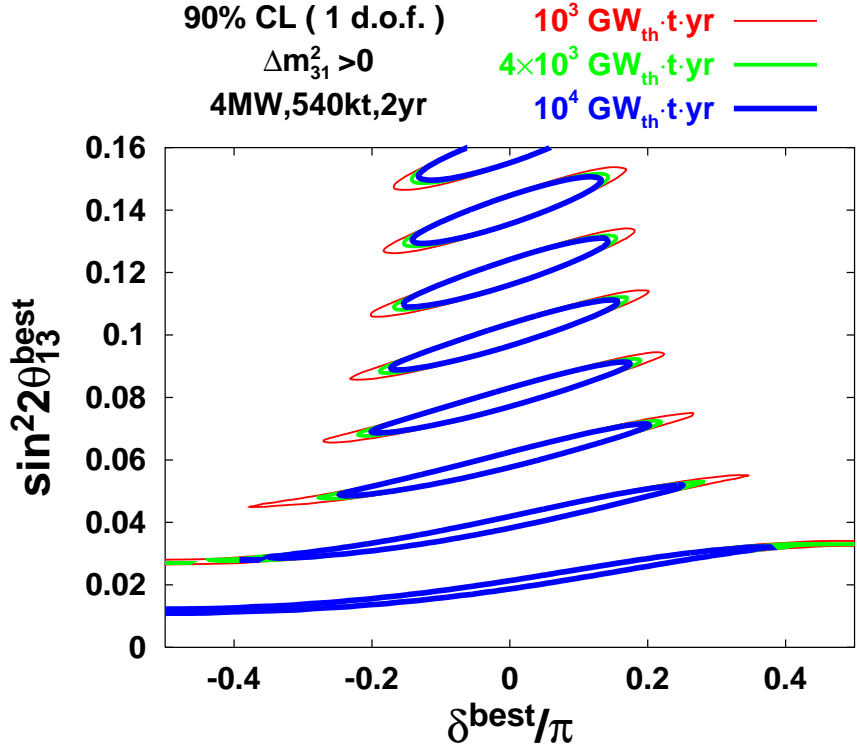,width=12.0cm}
\end{center}
\caption{
The contours are plotted for eight assumed values of 
$\sin^2{2\theta_{13}}$ which range from 0.02 to 0.16 
to indicate the regions consistent with the hypothesis 
$\delta = 0$ at 90\%~CL by the reactor-LBL combined measurement.
If an experimental best fit point falls into outside the envelope of 
those regions, it gives an evidence for leptonic CP violation at 
90\%~CL. 
The thin-solid (red), solid (green), and thick-solid (blue) 
lines are for 
$10^3$, $4\times 10^3$, and $10^4$ 
GW$_{th}$$\cdot$ton$\cdot$year  
exposure of a reactor experiment, respectively, 
corresponding to about 0.5, 2, and 5~years exposure of 
100~ton detectors at the Kashiwazaki-Kariwa nuclear power plant.
For the JPARC-HK experiment, 2~years measurement with 
off-axis $2^\circ$ $\nu_{\mu}$ beam is assumed. 
The normal mass hierarchy, $\Delta m^2_{13} > 0$, is assumed.
}
\label{fig:rCP_HK2}
\end{figure}

\begin{figure}[h]
\begin{center}
\epsfig{figure=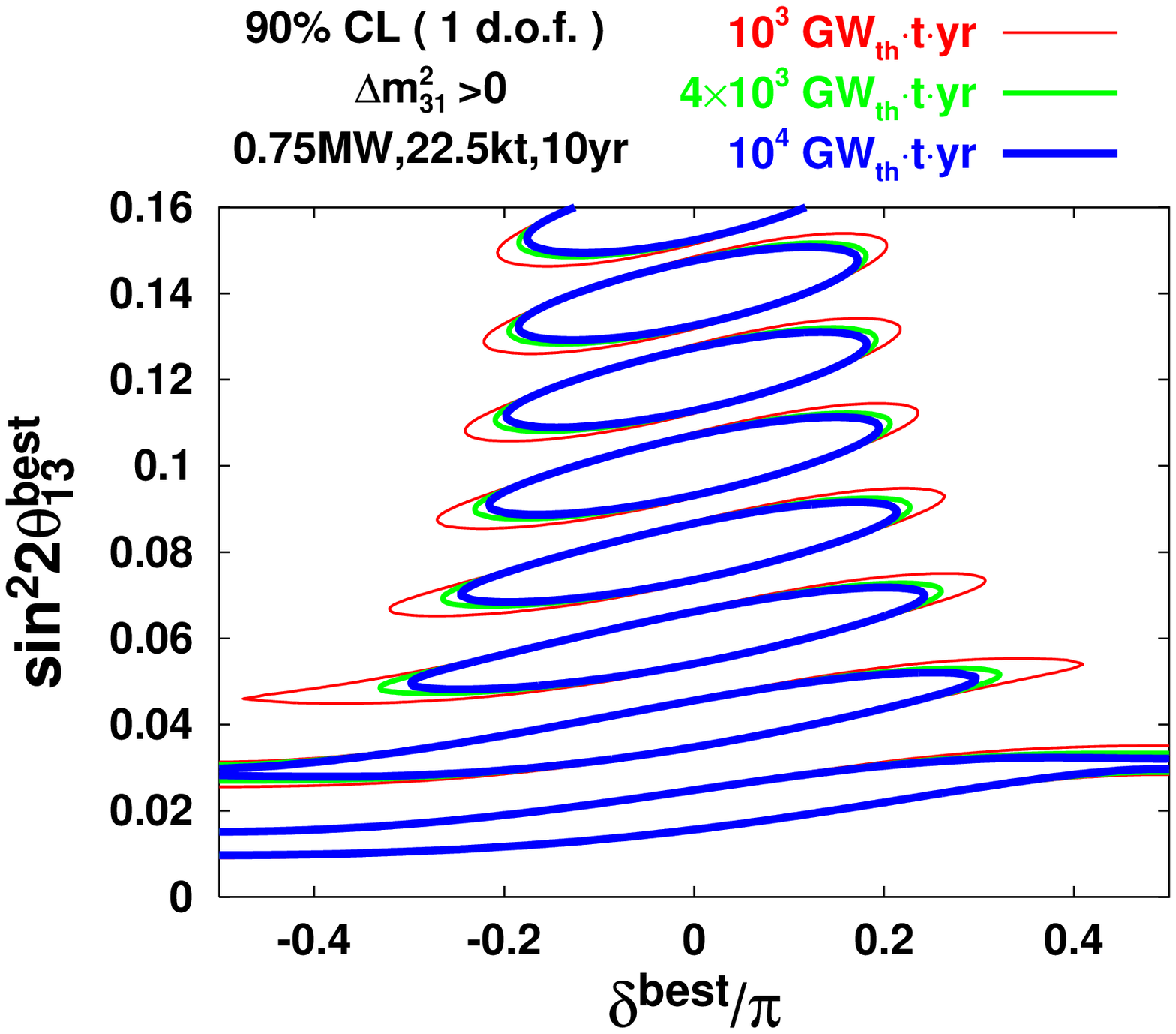,width=12.0cm}
\end{center}
\caption{
The same as in Fig:~\ref{fig:rCP_HK2} but with 10 years running 
of JPARC-SK experiment with the fiducial volume of 2.25 kton and 
the beam power of 0.75 MW. 
}
\label{fig:rCP_SK10}
\end{figure}

We take the following procedure in our analysis.
We pick up a point in the two-dimensional parameter space 
spanned by $\delta^{best}$ and $\sin^2{2\theta_{13}^{best}}$ 
and make the hypothesis test on whether the point is consistent 
with CP conservation within 90\%~CL. For this purpose, 
we use the projected $\Delta \chi^2$ onto one-dimensional 
$\delta$ space \cite{MS03}.
Then, a collection of points in the parameter space which 
are consistent with CP conservation form a region surrounded 
by a contour in 
$\delta^{best}$-$\sin^2{2\theta_{13}^{best}}$ space, 
as will be shown in Figs:~\ref{fig:rCP_HK2} and \ref{fig:rCP_SK10}.
Or conversely, if an experimental best fit point falls into 
outside the envelope of those regions, it gives an indication 
for leptonic CP violation
because it is inconsistent with the hypothesis $\delta = 0$
at 90 \% CL. 
We take the neutrino mixing parameters which correspond to 
the LMA-I solar neutrino solution.

In Fig:~\ref{fig:rCP_HK2}, we present the results of the CP sensitivity
analyses by assuming neutrino-mode appearance measurement at
JPARC-HK for 2 years and reactor measurement of $10^3 - 10^4$
GW$_{th}$$\cdot\mbox{ton}\cdot\mbox{year}$. 
The fiducial volume of HK is taken as 540 kton and 
the upgraded beam power of 4 MW is assumed. 
In Fig:~\ref{fig:rCP_SK10}, we give the results of similar 
analysis by using 10 years running of SK with fiducial volume 
22.5 kton and beam power 0.75 MW. 
We observe from Fig:~\ref{fig:rCP_HK2} that there is a chance for 
reactor-LBL combined experiment of seeing an indication of CP 
violation for relatively large $\theta_{13}^{best}$, 
$\sin^2{2\theta_{13}^{best}} \geq 0.03$ at 90 \% CL.
Moreover, it is truely remarkable that the sensitivity to 
CP violation is not lost and remains for 
$\sin^2{2\theta_{13}^{best}} \geq 0.04$ with SK, though 
data taking by SK for 10 years may be painful. 
Without HK, the reactor-SK combined measurement might 
be the unique way of detecting leptonic CP violation.

The sign of $\Delta m^2_{13}$ is taken to be positive in 
Figs:~\ref{fig:rCP_HK2} and \ref{fig:rCP_SK10},
which corresponds to the normal mass hierarchy.
If we flip the sign of $\Delta m^2_{31}$ 
(the case of inverted mass hierarchy) 
we obtain almost identical CP sensitivity contours.
Then, it looks like that the sign of $\Delta m^2_{13}$ 
does not produce any serious problems. 
Unfortunately, it is not true. 
If we do not know the sign, there arises a severe limitation 
on our ability of detecting CP violation by the reactor-LBL 
combined method. In this case, one has to allow the possibility 
that we take a wrong sign in analyzing the reactor-LBL combined 
measurement. Then, one can show that about half of the sensitivity 
region is lost \cite{MS03}. 
This is nothing but the problem of parameter degeneracy 
due to unknown sign of $\Delta m^2_{13}$, the two-fold ambiguity 
first noticed in \cite{MNjhep}.
Therefore, determination of the sign of  $\Delta m^2_{13}$ 
must be done prior to the reactor-LBL combined measurement of CP 
violation. This is a part (only a part) of the reasons why 
I enphasized  the importance of measuring the sign of 
$\Delta m^2_{13}$ in my contribution in the panel discussion 
\cite{minakata}.
See Ref.~\cite{MS03} for details of the treatment of background 
and the complexity which arises due to the unknown sign.

\subsection{BNL Strategy; A Brief Review}

An interesting new idea was developed by people in Brookhaven National 
Laboratory which is described in \cite{BNL}. Let me briefly review it. 
The basic idea is as follows. 
As I described earlier, the conventional way of detecting CP violation 
is to compare $P(\nu_{\mu} \rightarrow \nu_{e})$ with 
$P(\bar{\nu}_{\mu} \rightarrow \bar{\nu}_{e})$ at around the 
first oscillation maximum, 
$\frac{\Delta m^2_{13}L}{4E}=\frac{\pi}{2}$. 
Instead of following the canonical path, one can think of a new 
possibility that if one could observe a full neutrino oscillation pattern 
the effect of $\delta$ should manifest itself. 
To incorporate this idea into the experiment it must be able to see not 
only the first but also the second and the third oscillation maxima. 
Then, one need to use high energy beam and longer baseline so that 
the oscillatory pattern is not to be buried into the Fermi motion. 
See Fig.~3 of the first reference in \cite{BNL}.
The design proposed by BNL group employs the wide-band beam 
from upgraded AGS to 1 MW whose energy specrum extend to $\sim$ 
6 GeV with baseline distance of 2540 km. 
The beam spread covers the first, the second and the third 
oscillation maxima which are located at about 5, 1.6, and 1 GeV, 
respectively.

The strategy and the resultant proposal seem to deserve closer 
attention assuming that severer background issue in water Cherenkov 
detector at higher energies is overcome. 
The group did a detailed simulations and estimated the sensitivities 
of various observables. So you can judge by yourself by looking 
into \cite{BNL}. Yet, let me make a few remarks,  
some good ones first and then a sour one next. 
Good ones are that there is a number of favorable features in the 
BNL strategy. Of course, it is nice to explore the neutrino oscillation 
pattern which has never been done in a clear way.\footnote{
%%%%%%%%%%%%%%%%%%%%%%%% footnote %%%%%%%%%%%%%%%%%%%%%%%%%%%%
An evidence for a dip at the first oscillation maximum 
in the ratio of data to Monte Carlo prediction in atmospheric 
neutrino observation has recently been reported by 
the Supar-Kamiokande group \cite{ishitsuka}, the first clear 
demonstration of the oscillatory behavior.
This is the comment inserted to this article in the last minute 
but I feel I should do it because it is so important. 
}
Despite the usage of a baseline longer than 2000 km its sensitivity 
goes down to rather small value of $\theta_{13}$, 
$\sin^2{2\theta_{13}} \simeq 0.005-0.04$ at 90 \% CL 
depending upon $\delta$ and the sign of $\Delta m^2_{13}$. 
In fact, the long baseline is advantagious for smaller 
values of $\Delta m^2_{13}$, and its sensitivity may extends 
even to the solar $\Delta m^2_{12}$.

My sour comment is on the sensitivity of detecting CP violation to be 
achieved by the concrete design of the experiment proposed by the BNL 
group. As indicated in Fig.~10 in the second reference in \cite{BNL}, 
the sensitivity of $\delta$ determination is limited to 20-30 degree at 
$1 \sigma$ CL for $\sin^2{2 \theta_{13}}=0.04$. It may be compared 
with the similar accuracy claimed by the JPARC-HyperKamiokande group 
but at $3 \sigma$ CL \cite{JHF}.
I want to note that it is not completely fair to compare the 
sensitivities claimed by the both groups in an equal footing. 
LOI of the JPARC-HK experiment assumes 4 MW as the beam power, 
whereas BNL proposal takes 1 MW.
But even after taking account of the difference in beam intensity, 
it appears to me that the much less sensitivity in CP detection in 
the BNL proposal arises because of limited statistics due to long 
baseline distance, which results in a factor of $\sim$ 100 less 
beam intensity than that of JPARC-HK per MW.\footnote{
%%%%%%%%%%%%%%%%%% footnote %%%%%%%%%%%%%%%%%%%%
This interpretation might be too naive if the claim that the 
sensitivity is approximately L-independent up to $\sim$ 4000 km 
\cite{marciano} is true.
}

Of course, it is not quite meaningful to compare the sensitivities
of the two experiments which employs different detection principles 
and differs in background and detector systematics.
The real question is if there are ways to improve the 
sensitivity along the line of the BNL strategy. 
I emphasize that if my prejudice that the statistics is of 
key issue in CP measurement is right, 
an L/E scaled realization of the BNL strategy by using the 
NuMI beam at Fermilab is worth to consider.

\subsection{A Low-Energy Realization of BNL Strategy}

Let me now address the last topics of this talk, a possible 
low-energy realization of the BNL strategy. 
My motivation is as follows:
the JPARC-SK program is now funded. 
(It was not quite decided at the time of the workshop, 
but it was likely as I indicated in my contribution at the 
panel discussion.)
Therefore, we will definitely have a neutrino superbeam in 
Japan in 2008 or so. Then, the natural question is 
``are there any possibilities of realizing the BNL strategy by 
utilizing the beam?''.

We cannot think of high energy beam, 
like the one in the BNL proposal, in the JPARC project, 
because the JPARC neutrino beam is an off-axis beam 
which will be tuned to $\sim$ 600 MeV, 
the first oscillation maximum at SK.
With baseline distance of 300 km the second and the third 
maxima are buried into the Fermi motion anyway. 
Therefore, the only possibility we can think of is the second 
detector specialized to measure neutrino oscillation at around 
the second oscillation maximum.\footnote{
%%%%%%%%%%%%% footnote %%%%%%%%%%%%%%%
This is different but reminiscent of the two detector proposal 
\cite{MNplb97} in its basic idea of detecting CP violation 
by observing neutrino oscillation at two different values of 
relative phase between two flavor states. 
}
Clearly, we need three-times longer distance to have 
$\frac{\Delta m^2_{13} L}{4E}=\frac{3 \pi}{2}$. 
It requires $L \simeq 900$ km from JPARC in Tokai village 
toward the direction to Kamioka; Korea!
\footnote{
%%%%%%%%%%%%  \footnote %%%%%%%%%%%%%
I am aware that the distance between Tokyo and Seoul is longer than 
900 km. It is more like 1100 km. Then, a question arises as to if our 
strategy of measurement at the first-second oscillation maxima works. 
My answer is as follows. Suppose that 
$\Delta m^2_{13}=2 \times 10^{-3}$ eV$^2$ 
as suggested by the most recent atmospheric neutrino analysis by 
SK group \cite{hayato}. 
Then, it may not be optimal to tune the beam energy to the 
first oscillation maximum at Kamioka because it is as low as 
480 MeV. It is quite conceivable to run at slightly higher 
energy and let me take it to 600 MeV. 
In this case, $\Delta_{13}^{Kamioka}=0.8 \pi$, slightly below the first 
oscillation maximum. 
Then, $\Delta_{13}^{Korea}=1.03 \times 3\pi$, which is almost the 
second oscillation maximum.
If $\Delta m^2_{13}=2.5 \times 10^{-3}$ eV$^2$, 
$\Delta_{13}^{Kamioka}=\pi$ and 
$\Delta_{13}^{Korea}=1.2 \times 3\pi$, $\sim$ 20 \% higher than the 
second oscillation maximum.
}

Let us examine the hypothetical experiment with one 
Hyper-Kamiokande (HK) in Kamioka  and the second HK in 
somewhere in Korea, a bit more expensive one compared with the 
current JPARC-HK project! 
We shoot neutrino superbeam from JPARC, I mean no 
antineutrino beam for this consideration, and detect them 
by the two HKs.

\begin{figure}[h]
\begin{center}
\epsfig{figure=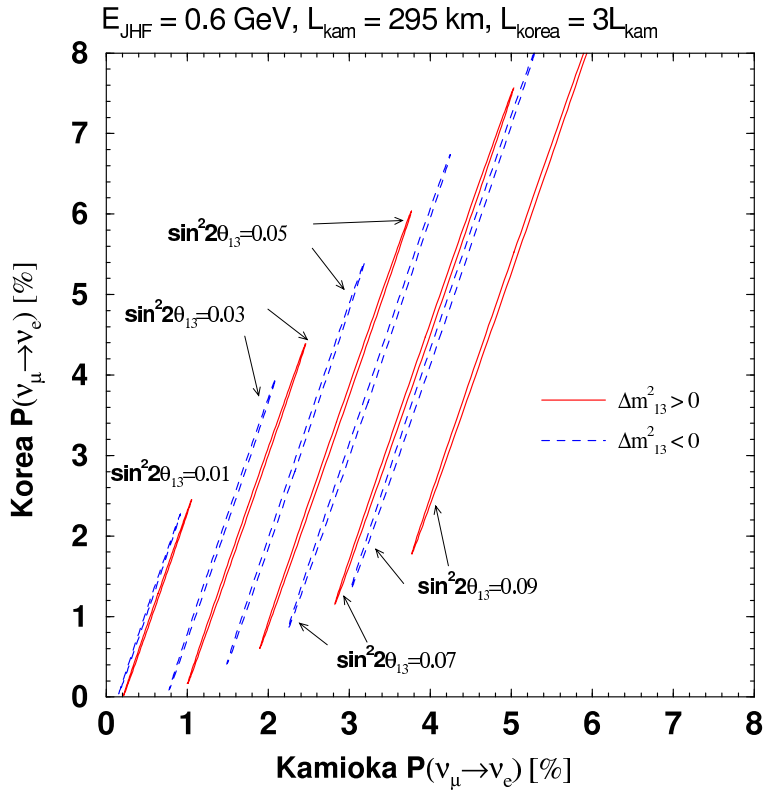,width=10.0cm}
\end{center}
\caption{
A Kamioka-Korea bi-probability plot is presented with neutrino beam energy 
600 MeV and the baseline lengths 
$L_{Kamioka}=295$ km and $L_{Korea}=3 L_{Kamioka}$.
The neutrino mixing parameters are taken as follows: 
$\Delta m^2_{13}=2.5 \times 10^{-3}$ eV$^2$ and 
$\theta_{23}=\pi/4$.
The LMA-I solar mixing parameters are taken as in \protect\cite{MNP3}.
}
\label{fig:jhf_korea}
\end{figure}

To reveal the qualitative features of the experiment I draw the 
bi-probability plot appropriate for the setting. The result 
is given in Fig:~\ref{fig:jhf_korea} in which 
the abscissa is taken to be the appearance probability 
$P=P(\nu_{\mu} \rightarrow \nu_{e})$
in Kamioka, and 
the ordinate is the appearance probability $P$ in Korea. 
The beam energy is assumed to be tuned to 600 MeV which 
is close to the first oscillation maximum at Kamioka for 
$\Delta m^2_{13}=2.5 \times 10^{-3}$ eV$^2$, 
and the monochromatic beam is assumed as a first approximation. 

A distinctive feature of the plot is that the slope of the shrunk 
ellipse is about -3 in contrast to the slope which is close to -1 
in the $P-\bar{P}$ plot. (See, e.g., \cite{KMN02}.) 
It can be easily understood as a consequence of the combined 
measurement at the first and the second oscillation maxima. 
To understand this point I use the expression of oscillation 
probability given in (\ref{Pmue}). 
One can easily show for given two measurement 
$P_1$ in Kamioka and $P_2$ in Korea 
\begin{eqnarray}
P_2 = - \frac{Y_2}{Y_1} P_1 + 
\frac{1}{Y_1} 
\left(X_1 Y_2 - X_2 Y_1 \right) s^2_{13} 
\end{eqnarray}
where $\pm$ index for the sign of $\Delta m^2_{13}$ is suppressed. 
Under vacuum oscillation approximation in (\ref{Y}), which should be a 
reasonable approximation in this setting, the slope of the 
shrunk ellipse can be estimated to be 
\begin{eqnarray}
\frac{Y_2}{Y_1} \simeq \frac{L_2}{L_1} = 3 
\end{eqnarray}
reflecting the ratio of the baseline distances corresponding 
to the first and the second oscillation maxima.

What does the steeper slope in the Kamioka-Korea bi-probability 
plot mean? It means that if the statistical error is comparable 
(assuming the systematic error is the same), the 
Korean detector is more important in determination of 
CP phase $\delta$. 
It is because the size of the ellipse is due to variation of  
$\delta$ and therefore the extension of the ellipse along the 
Korean axis implies the larger sensitivity for measuring $\delta$. 
Of course, the situation is not easy to realize, because 
we need to have 10 megaton HK in Korea to equalize numbers 
of events in both HK. So let me restrict myself into 
more ``conservative'' assumption of two identical HK 
in Kamioka and Korea. Then, the statistical error in 
Korean HK is worse by a factor of 3 and it exactly cancels 
the factor of 3 merit of the extended ellipse.

Thus, it looks like that the merit of Korean detector for 
measurment of $\delta$ is comparable with antineutrino 
measurement by HK in Kamioka. This is, of course, not quite 
true because one need 3 more years in the antineutrino channel 
to have equal number of events with the neutrino channel. 
Furthermore, the Kamioka-Korea detector complex 
can operate simultaneously with neutrino beam for 
$\sim$ 10 years with less relative systematic errors. 
Therefore, 8 years running of $\nu$ and $\bar{\nu}$ modes 
with one HK at Kamioka, admittedly very roughly speaking,
is equivalent to 2 years running of $\nu$ mode with the 
Kamioka-Korea two HK complex.
Unfortunately, the difficulty in determining the sign of 
$\Delta m^2_{13}$ still remains with us in this new type of measurement, 
as indicated in Fig:~\ref{fig:jhf_korea}.

I hope that I demonstrated the merit of possible Kamioka-Korean 
detector complex in the CP violation search.
At the very least, it may serve for a better understanding of the 
secret of power of the BNL strategy.
Moreover, I personally believe that this possibility is worth 
greater attention; 
if $\theta_{13}$ turns out to be smaller than the sensitivity 
limit of JPARC-SK search, 
then we might want to consider this possibility seriously.
In this case, we should think about operating both $\nu_{\mu}$ and 
$\bar{\nu}_{\mu}$ beams and optimization of relative time sharing 
is required. 
Of course, if one really thinks about the over-all merit 
of such project, one should also integrate various other capabilities 
including {\it in situ} determination of the sign of $\Delta m^2_{13}$.

I would like to mention here that the idea and interests for 
having underground detector in Korea has been described earlier 
\cite{korea}. 
I was delighted to learn during the workshop that the interest still 
continues to exist among people in Korea \cite{kim}.

\section{Concluding Remarks}

I have discussed some new ideas of how to explore leptonic CP violation 
which might prove profitable in certain circumstances. 
The reactor-LBL combined method is useful to start exploring CP violation 
simultaneously with $\nu$ mode operation of LBL 
without waiting for its $\bar{\nu}$ mode operation. 
While its sensitivity is limited and may not exceed 2 $\sigma$ CL, it can 
be the first indicator of CP violation if $\theta_{13}$ is relatively large, 
$\sin^2{2\theta_{13}} > 0.05$ or so. 
Any informations about CP phase $\delta$ is certainly of help to optimize 
the operation of $\nu$ and $\bar{\nu}$ mode operation of LBL experiments. 
Moreover, it may be the only way to explore leptonic CP violation if HK is 
not built, a pessimistic scenario. 

I also discussed extremely optimistic case of two HK complex, 
one at Kamioka and the second one in Korea. 
What is the real need for such an expensive option?
Well, we do not know the size of $\theta_{13}$, and the JPARC-SK 
experiment may fail to detect $\nu_{e}$ appearance, as cautiously 
remarked by Nishikawa \cite{nishikawa} in his comments in the 
panel discussion. 
If it turns out to be the case, we need much higher sensitivity for 
CP violation search. The sensitivity would be greatly increased if 
there is the second HK which can explore the region around the 
second oscillation maximum, as I suggested in an admittedly very 
rough treatment. I hope that it is a practical way to 
overcome the limitation of the sensitivity in CP violation search 
by the current design of the JPARC-HK project. It may worth serious 
attention if the $\nu_{e}$ appearance signal at the phase-I of the 
JPARC-SK experiment is vague at the verge of its sensitivity. 

In summary, I am under the strong feeling that the good-old strategy of 
detecting CP violation by measuring appearance probabilities 
$P$ and $\bar{P}$ by low-energy superbeam experiments is still the most 
promising one. Yet, we may need some other ideas for cases of 
unexpected situation such as tiny $\theta_{13}$, or financial difficulty 
which might prevent us from construction of larger detectors. 

The final goal of exploring CP violation 
may be resolving the parameter degeneracy completely, the topics 
that I failed to cover in my presentation. 
In short, it is the problem of multiple solutions for a given set of 
observable of $P$ and $\bar{P}$. It can be viewed as intrinsic degeneracy 
\cite{Burguet-C} in solution of a set of parameters ($\delta$, $\theta_{13}$), 
which is enriched by two kind of discrete degeneracies due to the sign 
of $\Delta m^2_{13}$ \cite{MNjhep} 
and the first-second octant ambiguity of $\theta_{23}$ \cite{octant}. 
See Ref.~\cite{MNP2} for global overview and a list of the references. 
At the present status of our understanding of the notorious problem, 
we may need extreme facility such as neutrino factory to solve it completely 
\cite{donini}. 
It is nice to see that an important step toward realization of the ultimate 
option is put forward in Europe \cite{spiro}.

\section{Acknowledgements}

I wish to express my deep gratitude to Milla Baldo Ceolin for her cordial 
invitation to the workshop, which will be remembered by the announcement 
of the joint Frejus-Gran Sasso project. 
I thank Hiroshi Nunokawa for many fruitful collaborations over the years 
and for kind preparation of Fig:~\ref{fig:jhf_korea}. 
The discussions and collaboration with Stephen Parke and 
Hiro Sugiyama were indispensable for this article. 
I also thank Bill Marciano for his critical comments.

\end{document}